\documentclass[final,5p,times,twocolumn]{elsarticle}%final,draft,preprint
\usepackage{booktabs}
\usepackage{tabular}
\usepackage{graphicx}
\usepackage{amssymb}
\usepackage{amsthm}
\usepackage[version=3]{mhchem}
\usepackage[switch]{lineno}
\usepackage{gensymb}
\DeclareMathOperator\erf{erf}
\usepackage{fixltx2e}
\usepackage[british]{babel}
\biboptions{sort&compress}
\journal{arXiv.org}

\begin{document}

\begin{frontmatter}
\cortext[cor]{Corresponding author}
 \author[]{E. N. Epie\corref{cor}}
 \ead{neepie@uh.edu}
\author[]{W. K. Chu}
\address{Physics Department and Texas Center for Superconductivity (T\ts{C}SUH), University of Houston, TX 77204, USA}

%\title{Using ion beam and thermal annealing to effectively create, control and monitor point defect concentration in ZnO.}
\title{ZnO Defect Modulation for Efficient Photocatalysis}
\newcommand{\tp}[1]{\textsuperscript{#1}}
\newcommand{\ts}[1]{\textsubscript{#1}}
\begin{abstract}
We have investigated the combined effect of low-energy self-implantation and thermal annealing on the near-surface defect concentration of ZnO single-crystals. Using ionoluminescence (IL), we demonstrate that a combination of low-energy low-fluence Zn and O ion implantation followed by annealing in Ar increases the near-surface point defect concentration in ZnO by two orders of magnitude. Point defects are known to increase the surface reactivity of ZnO, thereby improving the efficiency of ZnO photocatalytic processes such as bilirubin degredation. We hereby provide recommendations on how to improve the efficiency of ZnO-related photocatalytic processes.%Our results will be useful for the improvement of these and other related photo-catalytic processes.

\end{abstract}

\begin{keyword}

 Ion implantation\sep Point defects\sep Ionoluminescence\sep ZnO \sep Annealing \sep Catalysis

\PACS 78.60.Hk   \sep  61.72.uj \sep 61.72.-y \sep 71.55.Gs \sep 82.65.+r
%% Find PACS codes here: http://www.aip.org/pacs/pacs2010/individuals/pacs2010_regular_edition/index.html


\end{keyword}

\end{frontmatter}

%%\linenumbers

\newcommand{\tp}[1]{\textsubscript{#1}}
\newcommand{\ts}[1]{\textsuperscript{#1}}
\section{Introduction}
Toxic wastes substances, ranging from industrial effluents to metabolic bi-products, are matters of serious health and environmental concern. Photocatalysis is currently an important eco-friendly and cost effective method for such waste treatment.

\begin{figure}[htb!]
\centering 
\includegraphics[scale=0.25,keepaspectratio]{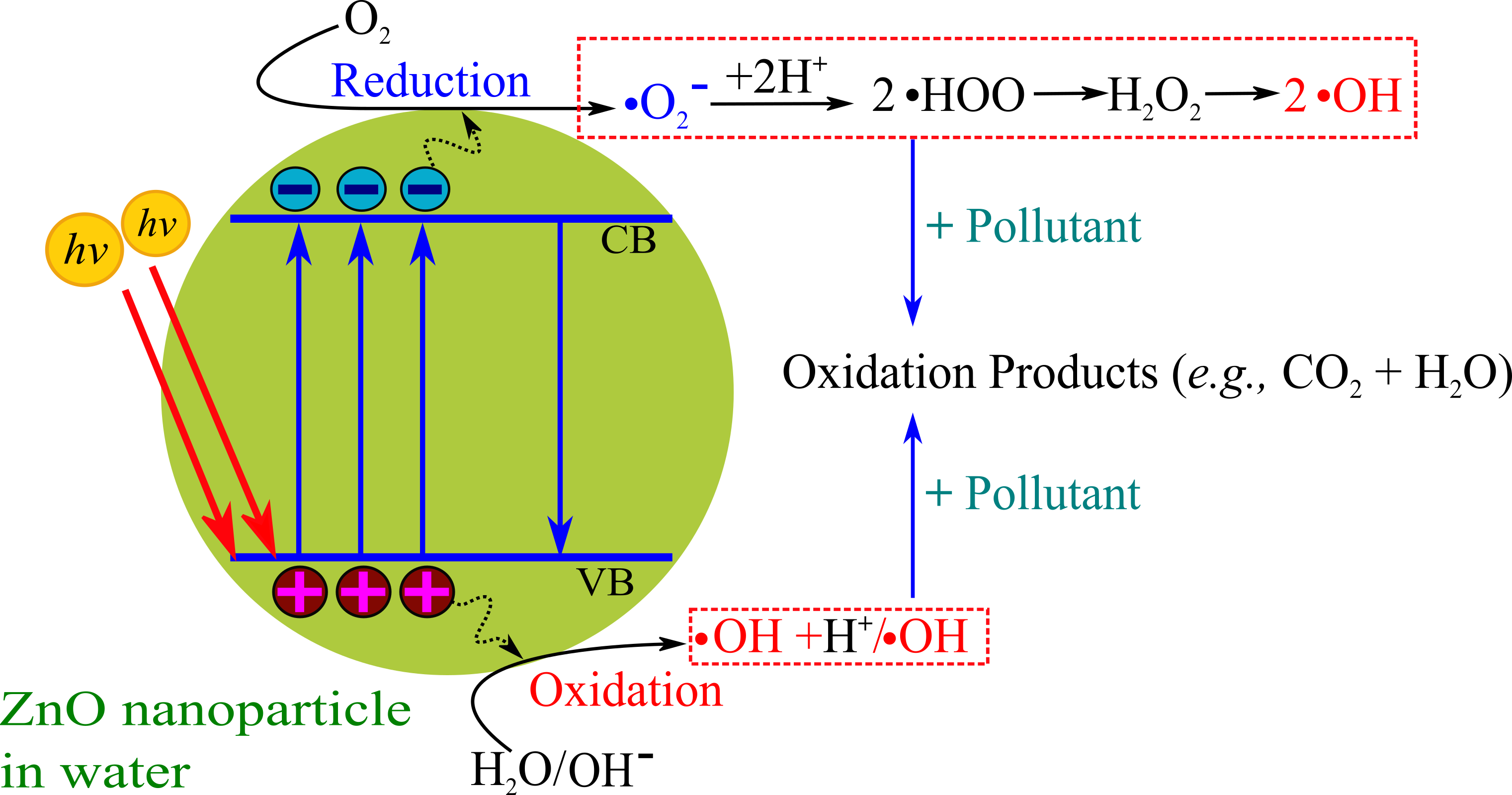}%width=\columnwidth
\caption{Schematic diagram of a typical photocatalytic process.}
\label{fig:ZnO_photocatalysis}
\end{figure}

%Photocatalysis is the use of light and a catalyst to bring about or accelerate a chemical transformation.
In an ideal photocatalytic process, organic pollutants are mineralized into \ce{CO2}, \ce{H2O} and mineral acids in the presence of a photocatalyst, typically a metallic oxide semiconductor, and an oxidizing species such as air. When semiconductors are exposed to light in solution, electrons-hole (e-h) pairs are generated. These e-h pairs aid the splitting of water into the hydroxyl (\ce{\cdot OH}) and superoxide (\ce{\cdot O2-}) radicals.  \ce{\cdot OH} is an extremely strong non-selective oxidant which leads to partial or complete mineralization of most organic pollutants. A typical photocatalytic process is illustrated in  Fig.\,\ref{fig:ZnO_photocatalysis}.

Commonly used metal oxide semiconductors for photocatalysis are titanium dioxide (\ce{TiO2}), ZnO, tungsten oxide (\ce{WO3}), strungtium titanate (\ce{SrTiO3}) and hematite ($\alpha$-\ce{Fe2O3}). Of these, ZnO is particularly attractive for photocatalysis due to a number of reasons. Firstly,  ZnO is biocompartible, biodegradable and non-toxic \cite{zho06}. Secondly, it has a high surface reactivity induced by native point defects \cite{ali10}. Thirdly, the ability to tune ZnO band gap $E_g$ through doping \cite{abd11} gives it a higher photoefficiency  in direct sunlight whose UV composition on Earth is significanlty low (4-7 \%). Lastly, ZnO is inexpensive and its nanostructures can be easily synthesized using a variety of methods. 

\begin{figure}[h!]
\centering 
\includegraphics[width=\columnwidth,keepaspectratio]{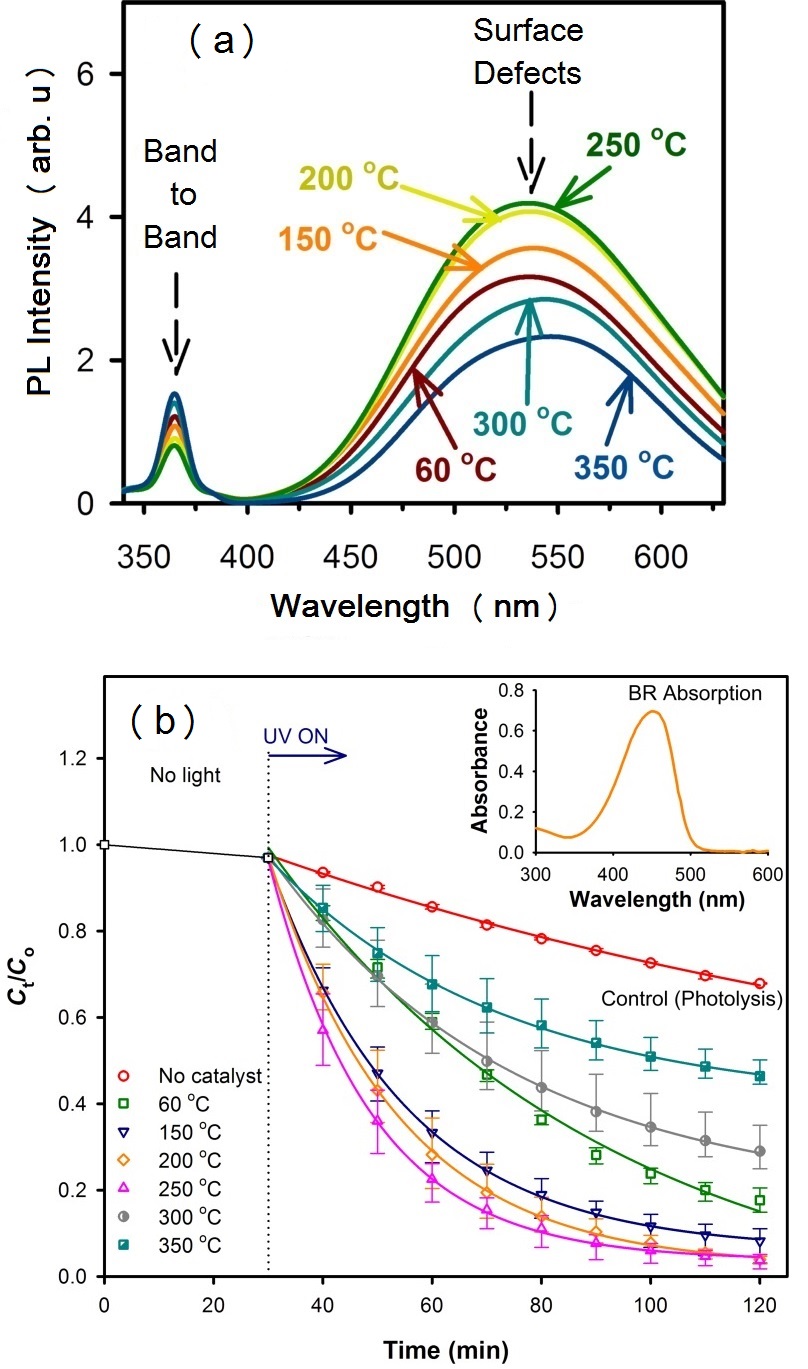}
\caption{(a) Room temperature PL spectra for ZnO nanoparticles annealed in air for 1 hr at various temperatures. (b) Bilirubin (BR) relative concentration ($C_t/C_0$) as a function of UV exposure time with and without (control) ZnO nanoparticles annealed in air at various temperatures. $C_t/C_0$ is measured from the BR absorption peak shown in the insert. Notice that the BR degradation rate is directly proportional to the near-surface defect concentration of ZnO nanoparticles. Addapted from \cite{bor13}.}
\label{fig:pl_ZnO_catalysis}
\end{figure}

Recently, Bora {\it et al.} \cite{bor13} investigated bilirubin (BR) degradation in the presence of ZnO nanoparticles and UV light.  They directly showed that ZnO nano-structures with a higher near-surface defect (particularly  oxygen vacancy) concentration greatly  accelerated bilirubin degredation in the presence of UV light, Fig.\,\ref{fig:pl_ZnO_catalysis}. Paradoxically, the primary method for oxygen vacancy ($\text{V}_{\text O}$) generation in ZnO for Bora and co-workers  was by thermal annealing of ZnO in air (\ce{O2}).  But, first-principle calculations show that the formation energy E\ts{f} of V\tp{O} in excess O is relatively high compared to that of V\tp{O} in excess Zn \cite{jan07}. Therefore, thermal annealing of ZnO in \ce{O2} is not the best approach for enhancing the near-surface $\text{V}_{\text O}$ concentration in ZnO for photocatalytic purposes.

In a recent study \cite{epi15b}, we demonstrated that a judicious combination of ion implantation and thermal annealing can better enhance the near surface defect concentration of ZnO.

We hereby investigate the combined effect of low-energy ion implantation and thermal annealing on the near-surface defect concentration of ZnO single-crystals. Our results show that the near-surface defect concentration in ZnO can be significantly enhanced through a careful combination of self-implantation and thermal annealing in Ar. Particularly, we show that  by simply annealing pristine ZnO in \ce{Ar} at 1000 \degree C for 1 hour,  the near surface defect concentration increases by a factor of 30. Annealing ZnO in \ce{O2} under similar conditions show a defect enhancement factor of only 7. These findings provide new perspectives on the efficiency enhancement of photocatalytic processes.  

\section{Experiment}
Double-face EPI-polished $ 5\times5\times0.5$ mm wafers of hydrothermally grown (MTI Corporation, Richmond, CA) c-plane ZnO single-crystals were used in this experiment. Sample purity was  greater than $99.99\,\%$ with minute traces of Mg $<0.0005\, \%$, Al $<0.003\, \%$, Si $<0.003\, \%$, Ti $< 0.001\, \%$, Cu $<0.003\, \%$, Fe $<0.005\, \%$, Ca $<0.0005\, \%$, and Ag $<0.0002\, \%$, according to the supplier.
 The Zn face of eight samples was modified by 60 keV self-implantation and thermal annealing  as described on Table \ref{tab:exp}. 
\begin{table}[h!]
\centering
\caption{Treatment conditions for ZnO single-crystals prior to IL analysis.}
\vspace{0.3cm}
\label{tab:exp}
\bgroup
\begin{tabular}{l | l | l}%{|l@{\hskip 0.25 in}|c|@{\hskip 0.25 in}c|}
\hline \hline
%\toprule
Sample name & 60 keV Implantation  & Annealing in 1 atm\\
%\midrule
\hline \hline
ZnO &none (virgin) & none \\
ZnO\tp{O\tp{2}} &none (virgin) & 1000 \degree C in O\tp{2}, 1 hr \\
ZnO\tp{Ar} & none (virgin)& 1000 \degree C in Ar, 1 hr\\
ZnO:8O & O\ts{-}/ $8 \times 10^{15}\ \text{cm}^{-2}$ & none \\
ZnO:8Zn & Zn\ts{-}/ $8\times 10^{15}\ \text{cm}^{-2}$ & none  \\
ZnO:8O\tp{Ar} & O\ts{-}/ $8 \times 10^{15}\ \text{cm}^{-2}$ & 1000 \degree C in Ar, 1 hr \\
ZnO:8Zn\tp{Ar} & Zn\ts{-}/ $8\times 10^{15}\ \text{cm}^{-2}$ &1000 \degree C in Ar, 1 hr \\
ZnO:16O\tp{Ar} & O\ts{-}/ $1.6 \times 10^{16}\ \text{cm}^{-2}$ & 1000 \degree C in Ar, 1 hr \\
ZnO:16Zn\tp{Ar} & Zn\ts{-}/ $1.6\times 10^{16}\ \text{cm}^{-2}$ &1000 \degree C in Ar, 1 hr \\
\bottomrule
%\hline
\end{tabular}
\egroup
\end{table}

Implantations were carried out in a vacuum chamber with pressure $P=10^{-7}$ Torr. The O\ts{-} and Zn\ts{-} beam current densities were 6 $\mu$A cm\ts{-2} and 0.24 $\mu$A cm\ts{-2}, respectively. Ion beams were generated by a source for negative ions by cesium sputtering (SNICS) \cite{mid83}. For all practical purposes, it does not matter whether implantation is made with positive or negative ions. To restore bulk crystallinity while simultaneously increasing the near-surface point defect concentration, samples were annealed under Ar and \ce{O2} at 1000 \degree C.  All samples were then irradiated by 2 MeV He\ts{+} ions and the resultant ionoluminescent (IL) spectra obtained. Details of the IL process are described in Ref. \cite{epi15b}.  IL spectra for all samples were then analysed.\\
The relative near-surface defect concentration for each sample was monitored using an enhancement parameter defined by
\begin{equation}\label{eq:def_il}
\eta =(\text{I}_{\text g-y}/\text{I}_{\text uv}),
\end{equation}
where $\text{I}_{\text g-y}$ and $\text{I}_{\text uv}$ are the green-yellow and UV luminescent band intensities, respectively. The smaller this ratio is, the better the crystalline quality \cite{wil10}. 

At high fluences ($\phi \geqslant10^{16}$ \ce{cm^{-2}}), sputtering, and ion-beam induced migration of atoms can significantly reduce the projected range $R_p$ and ion concentration. Thus, fluence-dependent depth distribution for implant species was estimated using the equation, \cite{wan13,gna08}:
\begin{equation}\label{eq:sput_fluene_dist}
G(x,\phi)=\frac{N}{2Y}\biggl[\erf\biggl(\frac{x-R_p+\phi Y/N}{\Delta R_p \sqrt{2}}\biggl)-\erf\biggl(\frac{x-R_p}{\Delta R_p \sqrt{2}}\biggl)\biggl]
\end{equation}
 where $x$ is the depth coordinate with respect to the instantaneous target (or substrate) surface, $N$ is the atomic density of the substrate, $Y$ is the sputtering yield, $\Delta R_p$ is the range straggling related to $R_p$ and ``$\erf(\,)$'' is the error function. All fitting parameters to Eq.\,\ref{eq:sput_fluene_dist} were obtained from computer simulations with the SRIM2008 \cite{zieg08} code.

\section{Results and Analysis}
\subsection{SRIM Analysis}
\begin{figure}[htb]
\centering 
\includegraphics[width=\columnwidth,keepaspectratio]{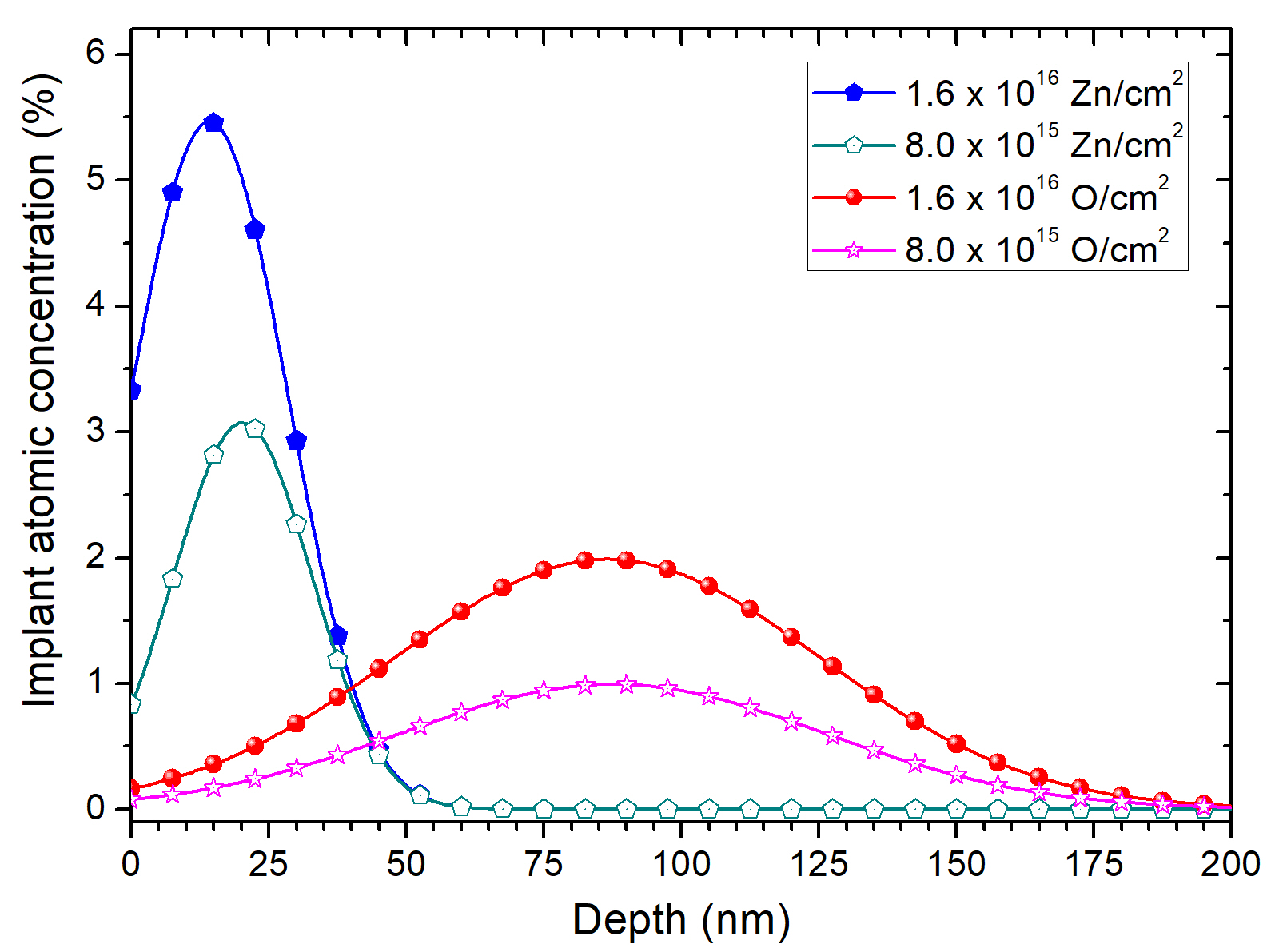}
\caption{Estimated 60 keV Zn and 60 keV O implant concentration profiles in ZnO based on Eq.\,\ref{eq:sput_fluene_dist} which appropriately accounts for target sputtering.}
\label{fig:Zn_O_dist_all}
\end{figure}

Fig.\,\ref{fig:Zn_O_dist_all} shows the fluence-dependent implant distribution based on Eq.\,\ref{eq:sput_fluene_dist}. While the projected range for O implants remain constantly at 87 nm, that  for Zn implant shifts towards the surface from 20 nm to 15 nm as fluence doubles from $8\times10^{15}\, \text{to}\, 1.6\times10^{16}\, \text{cm}^{-2}$. This is due to the relatively high sputtering rate of ZnO induced by Zn ions. In fact, under our  experimental conditions, the estimated sputtering yields due to Zn\ts{-} and O\ts{-} ions are 13 atoms/ion and 2 atoms/ion, respectively.  Although Zn and O implant concentrations roughly double as corresponding fluences double, Zn implants at high fluence are likely to reach the ZnO target surface earlier and anneal out faster than those at low fluence upon heat treatement. This is due to smaller instantaneous R\tp{p}, narrower distribution profile and higher mobility of $\text{Zn}_i$ compared to low fluence Zn and O implants.

\subsection{IL Analysis}
\begin{figure}[htb!]
\centering 
\includegraphics[width=\columnwidth,keepaspectratio]{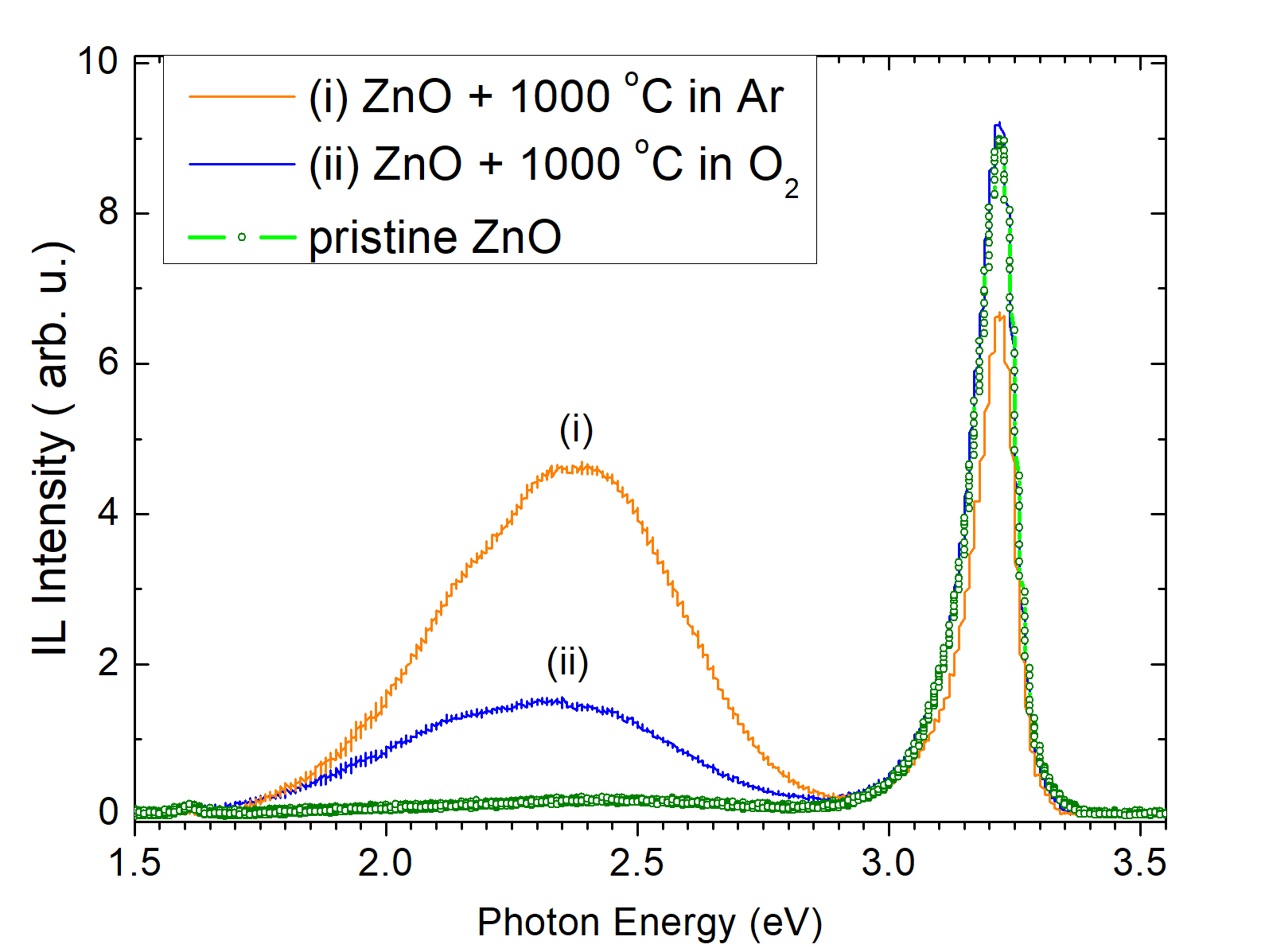}
\caption{IL spectra of pristine ZnO single-crystals annealed under \ce{O2} and Ar. For the Ar and \ce{O2} annealed samples, I\tp{g-y}(ZnO\tp{Ar}) = 3 I\tp{g-y}(ZnO\tp{O\tp{2}}).}
\label{fig:v3}
\end{figure}
Fig. \ref{fig:v3} compares the IL spectrum of pristine ZnO  with those of ZnO annealed under ambient Ar and O\tp{2} at 1000 \degree C for 1 hr. All samples display three peaks, characteristic of ZnO: a sharp UV emission peak centered around 3.2 eV, a broad green-yellow emission peak centered around 2.4 eV, and a small red emission peak centered around 1.6 eV. We observe significant enhancement of I\tp{g-y} for the annealed samples compared to pristine ZnO. A low I\tp{g-y} and correspondingly high I\tp{uv} in the pristine sample signifies very small near-surface defect concentration in agreement with the high crystalline quality of this pristine sample. On the other hand, the high I\tp{g-y} for annealed ZnO samples indicate an increase in their near-surface defect concentration. Particularly, I\tp{g-y}(ZnO\tp{Ar}) = 3 I\tp{g-y}(ZnO\tp{O\tp{2}}) suggesting that the near surface defect concentration in ZnO\tp{Ar} is at least 3 times that in ZnO\tp{O\tp{2}} ({\it i.e.}, $\eta_{Ar}\geqslant 3\, \eta_{\ce{O2}}$).

Previously we demonstrated that the 2.4 eV peak is mostly related to near surface V\tp{O} and presumably $\text{Zn}_i$ concentrations \cite{epi15b}. Although both defects have almost the same defect-formation energy E\ts{f}, according to first principle calculations \cite{jan07}, the defect-migration-barrier potential for Zn\tp{i} is lower than that for V\tp{O}. Thus, Zn\tp{i} are more mobile than V\tp{O}. This implies that both defects are created with equal ease but Zn\tp{i} are likely to anneal faster. Therefore, we deduce that the the luminescent effect of Zn\tp{i} on the 2.4 eV peak is minimal.

During ZnO annealing under ambient \ce{O2}, the rate of near-surface V\tp{O} annihilation is enhanced. This is because ambient \ce{O2} diffuses into the ZnO bulk recombining with some V\tp{O}. This explains why I\tp{g-y}(ZnO\tp{Ar}) = 3 I\tp{g-y}(ZnO\tp{O\tp{2}}).  
\begin{figure}[h!]
\centering 
\includegraphics[width=\columnwidth,keepaspectratio]{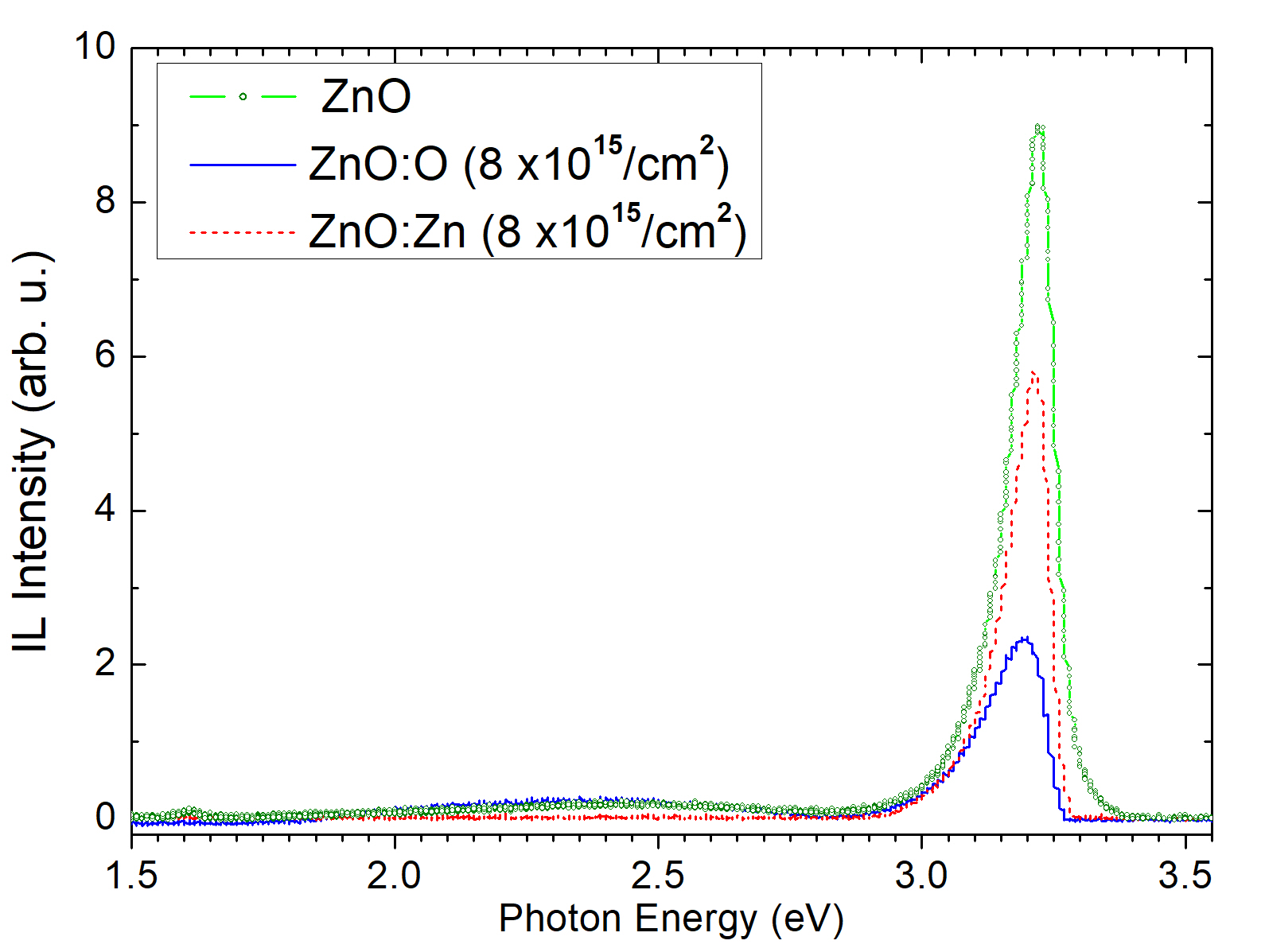}
\caption{IL spectra of ZnO single-crystal; as received and 60 keV fixed fluence ($8\times 10^{15}\ \text{cm}^{-2}$) O\ts{-} and Zn\ts{-} implanted. There is a significant drop in the UV emission band but no enhancement in green-yellow emission band.}
\label{fig:imp3}
\end{figure}

Ion implantation is another effective method of introducing point defects in ZnO. Fig.\,\ref{fig:imp3} compares the IL spectra of \ce{Zn^-} and \ce{O^-} implanted ZnO with that of pristine ZnO single-crystal. A drop in I\tp{uv} for as-implanted samples clearly indicates the presence of point defects resulting from radiation damage. Paradoxically, there is no corresponding enhancement of I\tp{g-y}. This is due to the intensification of competitive non-radiation pathways induced by stress fields generated in the implantation layer. Thermal annealing is an effective way to relax bonds thereby suppressing these non-radiation pathways.

\begin{figure}[h!]
\centering 
\includegraphics[width=\columnwidth,keepaspectratio]{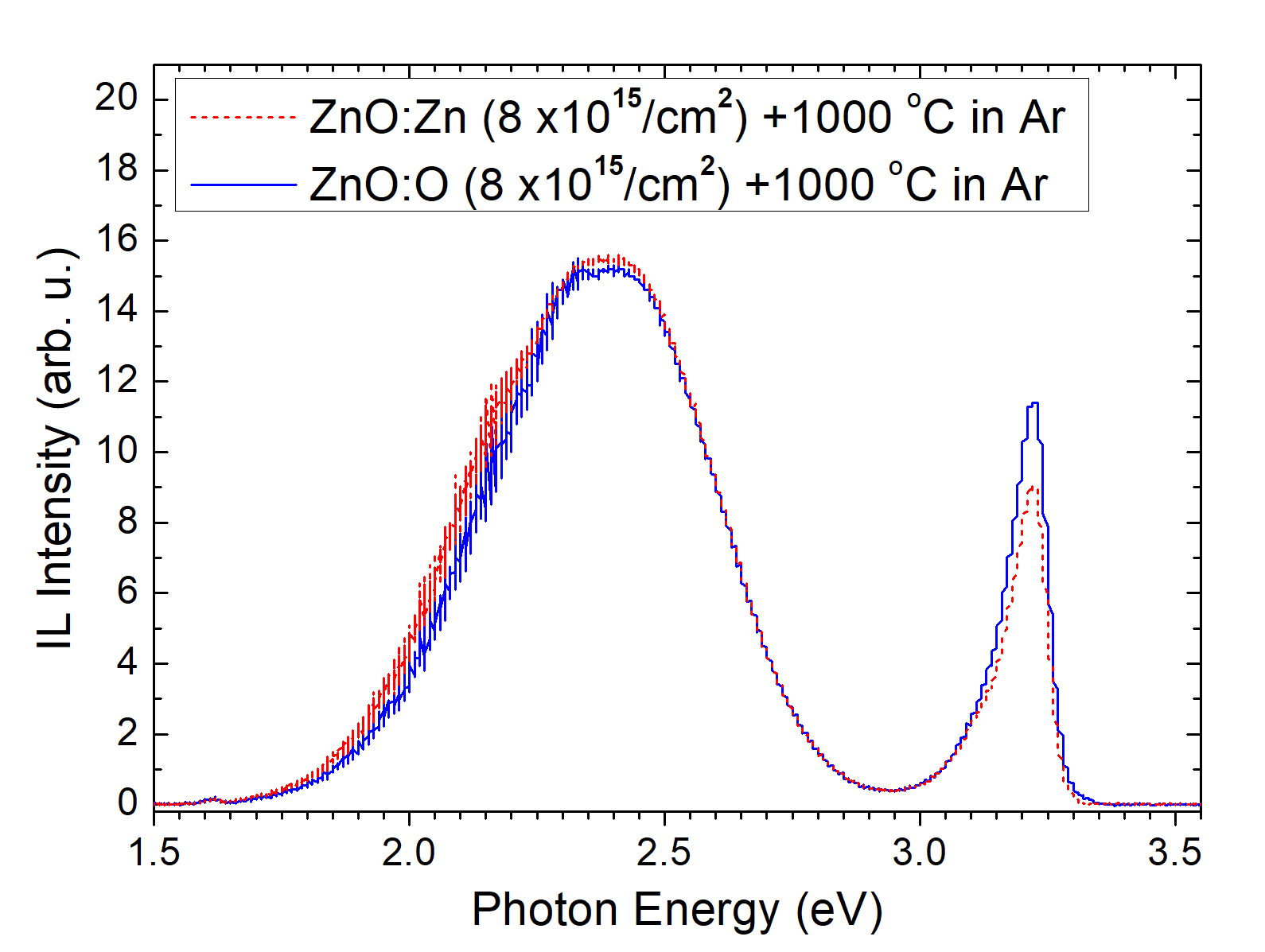}
\caption{IL spectra of annealed fixed fluence ($8\times 10^{15}\ \text{cm}^{-2}$) 60 keV O\ts{-} and Zn\ts{-} implanted ZnO single-crystals. The green-yellow luminescence intensity in both samples is greatly enhanced in equal proportions.}
\label{fig:8e15ann}
\end{figure}
Fig. \ref{fig:8e15ann} shows the effect of thermal annealing on the IL spectra of self-implanted ZnO single-crystals. We observe significant equal enhancement of I\tp{g-y} for both the ZnO:8O\tp{Ar} and ZnO:8Zn\tp{Ar} samples. In fact, I\tp{g-y} for both samples is more than 3 times that for the ZnO\tp{Ar} sample and more than 9 times that for the \ce{O2} annealed ZnO shown in Fig.\,\ref{fig:v3}. On the other hand, I\tp{uv} for both ZnO:8O\tp{Ar} and ZnO:8Zn\tp{Ar} samples is slightly less than that for pristine ZnO. Using Eq.\,\ref{eq:def_il}, we estimate the relative defect concentration ratios $\eta/\eta_o$  for  ZnO:8O\tp{Ar} and ZnO:8Zn\tp{Ar} samples at 56 and 72 respectively; $\eta_o$ being the near-surface defect parameter for pristine ZnO.  
\begin{figure}[h!]
\centering 
\includegraphics[width=\columnwidth,keepaspectratio]{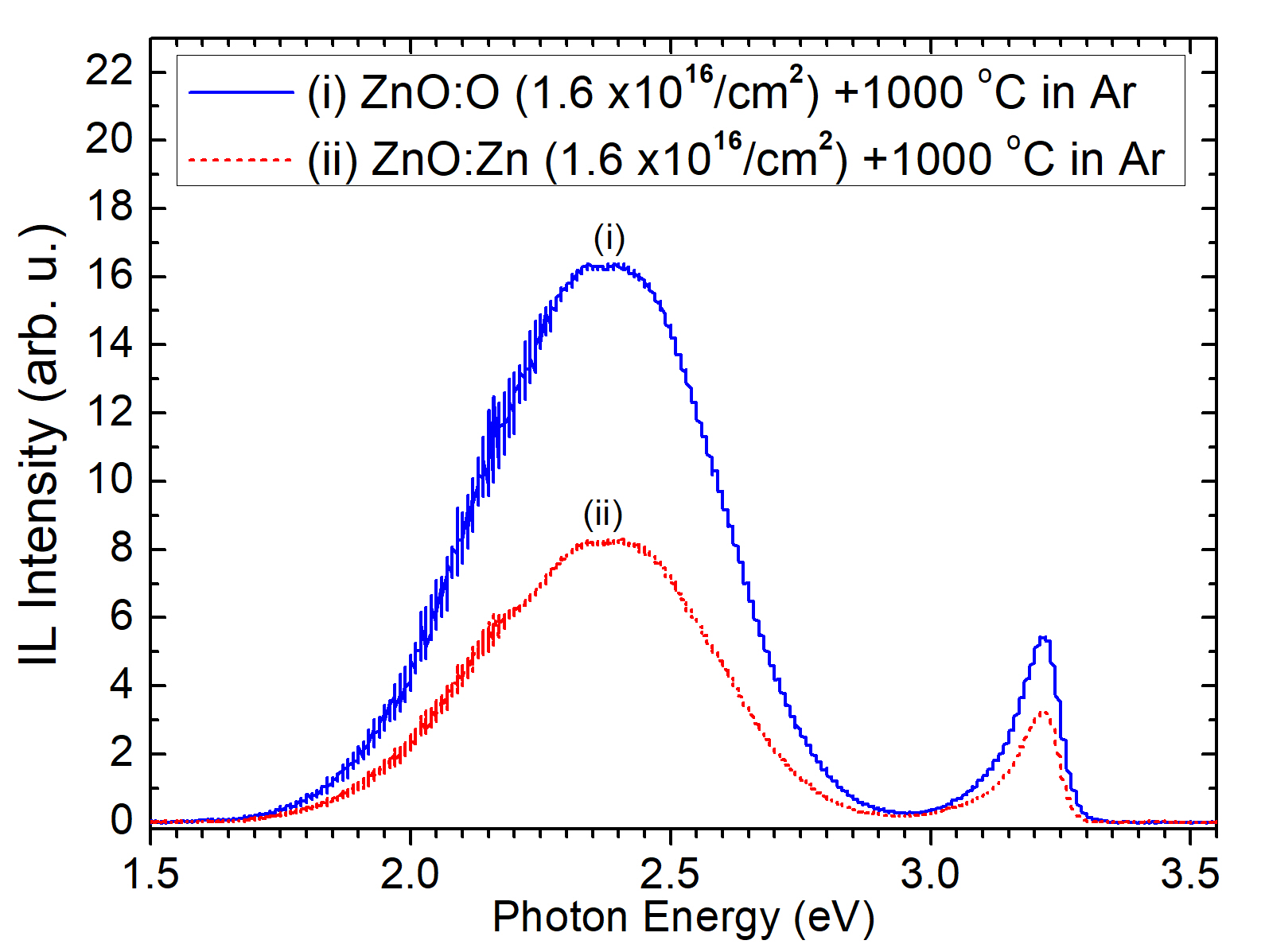}
\caption{IL spectra of annealed fixed fluence ($1.6\times 10^{16}\ \text{cm}^{-2}$) 60 keV \ce{O^-} and \ce{Zn^-} implanted ZnO single-crystals.}
\label{fig:1.6e16ann}
\end{figure}

Fig.\,\ref{fig:1.6e16ann} shows the effect of doubling Zn and O implant fluence followed by thermal annealing. We observe that by doubling the implantation fluence to  $1.6 \times 10^{16}\ \text{cm}^{-2}$, I\tp{g-y}(ZnO:16O\tp{Ar}) remains high while I\tp{g-y}(ZnO:16Zn\tp{Ar}) drops relative to  ZnO:8Zn\tp{Ar}. However, these changes are accompanied by corresponding decrease in both I\tp{uv}. Using Eq.\,\ref{eq:def_il}, the relative defect concentrations for ZnO:16O\tp{Ar} and ZnO:16Zn\tp{Ar} are respectively 125 and 104. These values are almost double those obtained for samples  ZnO:8O\tp{Ar} and ZnO:8Zn\tp{Ar}.

\section{Summary and Conclusions}
We have used a combination of thermal annealing, and low-energy self-implantation to modulate the near-surface native-point-defect concentration of ZnO single-crystals. Using ion beam induced luminescence, we have estimated the near-surface defect enhancement factor of our modified samples. Our results are summarised in Table \ref{tab:res}. We observe that by simply annealing pristine ZnO single-crystals in \ce{O2} and Ar, their near surface defect concentration, relative to pristine ZnO, increase by factors of 7 and 30, respectively. Low-energy self-implantation followed by thermal annealing further increases the samples' near-surface defect concentration upto 125 times that of pristine ZnO. 

\begin{table}[h!]
\caption{Estimates of  relative near surface defect concentration ($\eta/\eta_o$) for modified ZnO samples based on Eq.\,\ref{eq:def_il} with $\eta_o$ as the near surface-defect parameter for pristine ZnO.}
\label{tab:res}
\vspace{0.3cm}
\bgroup
\centering
\begin{tabular}{l | c | c | c | c}%{|l@{\hskip 0.25 in}|c|@{\hskip 0.25 in}c|}
\hline\hline
Sample name & I\tp{g-y} & I\tp{uv}& $\eta=(\text{I}_{\text g-y}/\text{I}_{\text uv})$ & $\eta/\eta_o$\\
\hline\hline
ZnO & 0.216 & 9.011 & 0.024 & 1 \\
ZnO\tp{O\tp{2}} & 1.529 & 9.203 & 0.166 & 7 \\
ZnO\tp{Ar} & 4.674 & 6.672 & 0.701 & 30 \\
ZnO:8O &  0.261 & 2.362 & 0.110 & /\\
ZnO:8Zn & 0.060 & 5.814 & 0.010 & / \\
ZnO:8O\tp{Ar} &15.288 & 11.408 & 1.340 & 56\\
ZnO:8Zn\tp{Ar} & 15.612 & 9.014 & 1.732 & 72 \\
ZnO:16O\tp{Ar} & 16.373 & 5.444 & 3.008 & 125\\
ZnO:16Zn\tp{Ar} & 8.297 & 3.22 & 2.498 & 104\\
\hline
\end{tabular}
\egroup
\end{table}
Thus, we an effective practical method for enhancing the photocatalytic activity of ZnO nano-structures in large scale applications is to anneal them under ambient Ar rather than \ce{O2}. Furthermore, irradiating ZnO nano-structures with low-energy low-fluence \ce{O^-} or \ce{Zn^-} ions followed by thermal annealing in Ar will greatly enhance their photocatalytic efficiency by even higher others of magnitude.
\section{Acknowledgments}
This work was funded by the state of Texas through the Texas Center for Superconductivity at the University of Houston. 

\section*{References}
\bibliographystyle{model1-num-names}%{unsrt}
\bibliography{epiebib_apsuscf}
\end{document}